\documentclass{article}
\usepackage{iclr2027_conference,times}

\usepackage{amsmath,amsfonts,bm}

\def\eqref#1{equation~\ref{#1}}

\def\1{\bm{1}}

\DeclareMathAlphabet{\mathsfit}{\encodingdefault}{\sfdefault}{m}{sl}
\SetMathAlphabet{\mathsfit}{bold}{\encodingdefault}{\sfdefault}{bx}{n}

\usepackage{hyperref}
\usepackage{url}
\usepackage{booktabs}
\usepackage{multirow}
\usepackage{array}
\usepackage{tabularx}
\usepackage{graphicx}
\usepackage{xcolor}
\usepackage{enumitem}
\usepackage{xspace}
\usepackage{placeins}

\newcommand{\bench}{\textsc{VGBench}\xspace}
\newcommand{\mute}{\texttt{[Mute]}\xspace}

\iclrfinalcopy 

\title{Do Audio LLMs Listen Before They Act?\\
Diagnosing Acoustic-Context Gating in Voice Agents}

\author{Yanjie Zhang$^{1*\dagger}$, Nanchen Hu$^{1*}$ \& Yushi Sun$^{2\ddagger}$\\
$^{1}$HKUST, Hong Kong SAR, China\\
$^{2}$LIGHTSPEED, Shenzhen, China\\
\texttt{\{yzhangj,nhuab,ysunbp\}@connect.ust.hk}
}

\begin{document}

\maketitle
{\renewcommand\thefootnote{}\footnote{$^{*}$Equal contribution.\newline\hspace*{1.8em}$^{\dagger}$Work done during Yanjie's internship at Tencent LIGHTSPEED.\newline\hspace*{1.8em}$^{\ddagger}$Corresponding author.}\addtocounter{footnote}{-1}}

\begin{abstract}
Audio language models can recognize spoken commands and invoke tools, but an
agent must first decide whether the acoustic and conversational context warrants
action. We introduce \bench, a 1,018-item diagnostic benchmark for action-level
addressedness across side-talk, self-talk, and speaker-switch scenarios. Each
item uses a shared action space comprising silence, a tool call, and a
natural-language answer. Speaker-switch pairs hold the specified words fixed
while source, distance rendering, and a temporal boundary define a controlled
wearer-to-bystander shift. Six raw Audio LLMs and three training-free
adaptations often identify the target tool yet rarely withhold action under this
shift; the highest raw switch mute rate is 14\%. We then use \textsc{VoxGate} as
a post-training case study. Supervised training mutes 91.3\% of switched commands
while choosing the correct tool for all nearby wearer commands and text-only
controls. An exploratory GRPO stage has similar switch performance; side-talk
accuracy rises from 68.4\% to 70.9\%, and self-talk muting from 52.0\% to 60.0\%.
Factorized controls identify an independent source-change effect, while sensitivity to the far-field manipulation varies across acoustic renderings.
The benchmark therefore measures multi-cue acoustic-context gating rather than isolated speaker identity.
\end{abstract}

\section{Introduction}
\label{sec:intro}

Voice agents increasingly perform consequential actions: they schedule
appointments, send messages, place orders, and control physical systems. A
spoken instruction carries both textual content, which specifies an action, and
acoustic and conversational context, which indicates whether the utterance is
addressed to the assistant. A bystander can utter the same command as the user,
and a user can mention a command while thinking aloud. An agent that relies on
text alone may recognize the requested operation while making the wrong decision
to act.

Existing systems often place keyword spotting, device-directed speech
detection, or speaker verification before the agent
\citep{mallidi2018devicedirected,nam2026speakerllm}. End-to-end
Audio LLMs instead expose one policy that can remain silent, answer, or invoke a
tool. Existing evaluations usually pair well-formed requests with a response or
tool label, so high tool-selection accuracy does not show that acoustic context
controls execution. We ask a narrower action-level question: when the specified
words are held fixed but the trigger comes from a different acoustic source and
scene, does an Audio LLM act or remain silent?

We study this question with \bench, a 1,018-item diagnostic benchmark for
\emph{agentic addressedness}. It contains 395 side-talk recordings, 223
self-talk recordings, and 400 paired speaker-switch cases. Every item maps to a
shared action space comprising \mute, a tool call, and a natural-language
answer. Side-talk tests conversational attribution within one recording;
self-talk tests command-like monologues; speaker-switch reverses the target from
a tool call to \mute while preserving the specified context and trigger words.
The switch condition jointly changes trigger source, far-field rendering, and a
600~ms boundary. It therefore tests a controlled wearer-to-bystander
source-and-scene shift, not isolated speaker identity.

We evaluate six raw Audio LLMs and three training-free adaptations under the
same standardized action contract. The strongest raw switch mute rate is 14\%,
and the strongest training-free result is 6\%, even when target-tool selection
is high. Step-Audio-R1.1, for example, selects the target tool on 96\% and 97\%
of same-speaker and text-only controls, respectively, but mutes only 1\% of
switched commands. These results expose a gap between recognizing command
content and using acoustic-pragmatic evidence to control action.

We then use \textsc{VoxGate} as a post-training case study. Supervised
fine-tuning combines the training partition of \bench with answer, abstention,
tool-use, and translation examples from WearVox. It accounts for most of the
observed gating improvement, muting 91.3\% of switched commands while choosing
the correct tool for all nearby wearer commands and text-only controls. An
exploratory counterfactual-pair GRPO stage yields similar switch muting. In one
run, side-talk accuracy rises from 68.4\% to 70.9\%, self-talk muting from
52.0\% to 60.0\%, and WearVox overall from 72.14\% to 76.30\%. These differences
do not isolate the effect of pair-aware grouping.
Factorized controls show that source change affects action independently, while
far-field rendering mutes 60\% of same-speaker triggers, as the proximity rule
requires. We further use equal-budget cue-balanced SFT to test how these cues
affect action.

Our contributions are:
\begin{itemize}[leftmargin=1.2em]
\item \textbf{Action-level benchmark.} \bench evaluates side-talk, self-talk,
and text-matched source-and-scene shifts under one silence, tool, and answer
interface.
\item \textbf{Behavioral diagnosis.} Full-corpus evaluation shows that current
systems can recover target tools while failing to condition execution on
acoustic and conversational evidence.
\item \textbf{Post-training case study.} Joint supervised training shows that
the measured gate is learnable without collapsing the positive controls;
exploratory GRPO and factorized controls characterize the remaining gains and
cue dependence.
\end{itemize}

\section{Related Work}
\label{sec:related}

\paragraph{Addressedness and voice-agent evaluation.}
Voice-assistant pipelines traditionally treat ``was I spoken to?'' as a
separate detection problem, using keyword spotting, device-directed-utterance
detection, speaker verification, or additional signals such as gaze
\citep{mallidi2018devicedirected,siegert2022addressee,zhang2025lookandtalk,nam2026speakerllm}.
Recent benchmarks bring related decisions into richer agent settings. WearVox
uses 3,842 real egocentric multichannel recordings and includes both side-talk
rejection and tool calling~\citep{lin2026wearvox}. Audio2Tool evaluates spoken
tool use across direct, compositional, and acoustically mixed queries
\citep{pahwa2026audio2tool}. ProVoice-Bench studies when proactive voice agents
should intervene or remain dormant~\citep{xu2026provoice}. \bench complements
these resources with text-matched action contrasts: the same specified command
supports execution in one condition and silence in another. This design connects
acoustic-context use directly to action selection rather than treating
addressedness as a separate front-end score.

\paragraph{Acoustic and paralinguistic evidence in Audio LLMs.}
MMAU and MMAU-Pro cover broad audio understanding and reasoning
\citep{sakshi2025mmau,kumar2026mmau}. Other benchmarks study prosody and
phonology in MMSU~\citep{wang2026mmsu}, stress and intonation in
WildSpeech-Bench~\citep{zhang2025wildspeech}, speaker attributes in
SD-Eval~\citep{ao2024sdeval}, multi-speaker grounding in
MSU-Bench~\citep{sun2026msubench}, and audio-dependent questions in
AUDITA~\citep{kabir2026audita}. Closest to our diagnostic design, VoxParadox
constructs linguistic-acoustic conflicts to test whether a model uses
paralinguistic cues~\citep{pang2026voxparadox}. Its targets are perceptual
answers; \bench instead asks whether acoustic and pragmatic evidence changes the
choice to remain silent, answer, or invoke a tool.

\paragraph{Post-training and inference-time adaptation.}
GRPO was introduced in DeepSeekMath~\citep{shao2024deepseekmath} and is now used
for audio reasoning post-training
\citep{li2025rlsft,wen2025sari,zhang2025aqattrl}. Omni-R1 shows that text-only
fine-tuning can improve audio benchmarks~\citep{rouditchenko2025omnir1}, while
recent methods explicitly measure audio contribution or penalize late-stage loss
of audio attention~\citep{he2026audiocontribution,xiao2026mapo}. Training-free
systems instead structure inference through tool orchestration, acoustic DSP, or
chunked listening
\citep{wijngaard2025audiotoolagent,maben2025aura,xiong2025tws,chiang2025shanks}.
We evaluate representative inference-time adaptations as behavioral probes and
use post-training to test whether the \bench action boundary is learnable.

\section{\bench: Action-Level Addressedness Diagnostics}
\label{sec:bench}

\subsection{Task Formulation}
An agentic voice assistant receives audio $x$, a system prompt $p$, and available
tools $\mathcal{T}$. It selects one of three actions: \textsc{Mute},
\textsc{Tool}, or \textsc{Answer}. \textsc{Mute} returns the literal token
\mute and executes nothing; \textsc{Tool} emits one structured call; and
\textsc{Answer} returns natural language without invoking a tool. We call the
joint decision based on textual semantics, intended recipient, and acoustic
source \emph{agentic addressedness}. Unlike perceptual classification, this task
tests whether contextual evidence changes an executable decision.

\begin{figure*}[t]
    \centering
    \includegraphics[width=\textwidth]{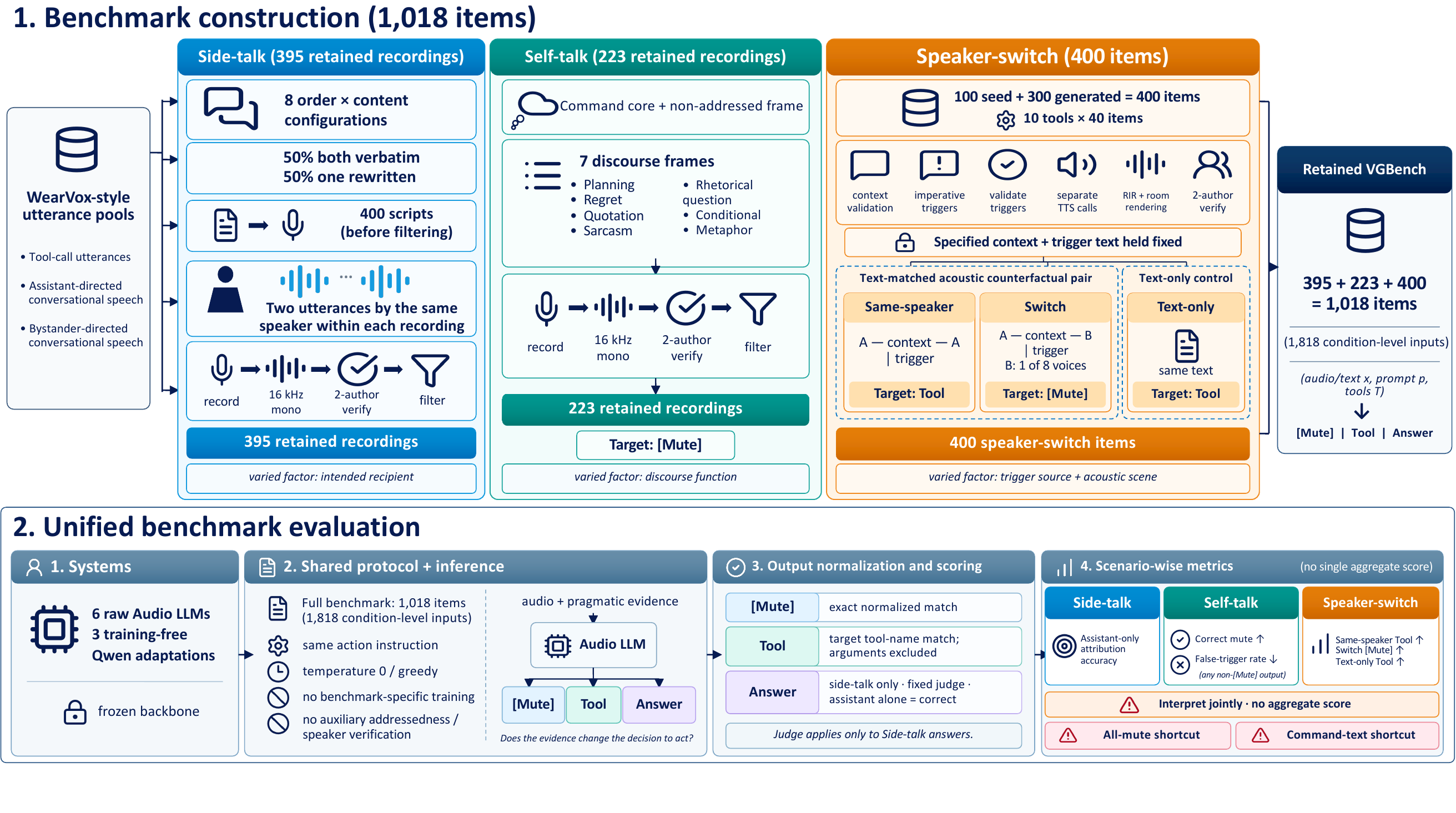}
    \caption{Overview of \bench construction and evaluation. The benchmark
    contains 1,018 items across side-talk, self-talk, and speaker-switch
    scenarios. Each speaker-switch item has same-speaker, switch, and text-only
    conditions. Metrics are interpreted jointly rather than collapsed into one
    aggregate score.}
    \label{fig:vgbench_pipeline}
\end{figure*}

\subsection{Diagnostic Design}
\bench uses three scenario families to separate complementary failures that are
usually collapsed into a single voice-assistant accuracy score. Side-talk asks
which utterance within a local exchange licenses action. Self-talk asks whether
command-like words are sufficient to trigger an action when the discourse frame
marks them as non-instructions. Speaker-switch pairs hold the specified words
fixed while reversing the target action under a controlled change in source and
scene. The positive controls attached to each family are essential for interpretation:
side-talk checks whether the model acts only on the addressed utterance,
self-talk tests whether it rejects lexical command shortcuts, and speaker-switch
requires action reversal while retaining tool recognition.

\paragraph{Side-talk.}
Each of 395 recordings contains two consecutive utterances from the same stored
speaker, one directed to the assistant and one to a nearby person, with balanced
order. Using one voice removes speaker identity as an explanation and tests
pragmatic attribution. The agent should act only on the assistant-directed
utterance. The scripts also balance tool-like and conversational content, so
utterance position or the presence of an obvious command is not by itself a
reliable decision rule.

\paragraph{Self-talk.}
The 223 recordings contain command-like monologues framed as planning, regret,
quotation, sarcasm, rhetorical questions, and related discourse forms. Every
item targets \mute even when the words contain a command that maps to an
available tool.

\paragraph{Speaker-switch.}
The 400 counterfactual pairs cover ten consequential tools. For these tool
triggers, we adopt a conservative wearable authorization rule:
$\mathrm{Tool}\iff\text{same authorized source}\land\text{near-field}$.
A different source or a far-field trigger targets \mute. In the
\emph{same-speaker} condition, source A (the session initiator)
speaks the context and near-field trigger, and the target is a tool call. In the
\emph{switch} condition, A speaks the context and
a bystander source B speaks the same trigger under fixed far-field rendering
after a 600~ms boundary, so the target is \mute. The \emph{text-only} condition
uses the same words and retains the tool target. The contrast holds specified
text fixed while jointly changing source, distance rendering, and temporal
boundary. A far-field trigger from A also targets \mute under this rule.

\subsection{Construction and Splits}
Side-talk and self-talk scripts are derived from tool-like,
assistant-directed, and bystander-directed seed pools. Their construction
balances utterance order and content type and places command cores in varied
discourse frames. Commissioned speakers recorded the scripts as natural speech.
Recordings are converted to 16~kHz mono PCM WAV and retained after two rounds of
annotation, target-label verification, and script-audio consistency checks,
yielding 395 side-talk and 223 self-talk recordings.

Speaker-switch uses 100 seed cases and 300 generated cases, balanced across ten
consequential tools with 40 cases per tool. Context and trigger segments are
synthesized separately using eight distinct voices to simulate speaker changes.
The same-speaker condition uses source A for both segments. The switch condition
uses source A for context, inserts 600~ms of silence, and renders the trigger from
source B with a fixed far-field room transform. Segments are normalized to
$-20$~dBFS with 5~ms fades before 16~kHz mono export. The manifest records voice,
duration, normalization, peak, boundary, and room configuration. These controls
make the aggregate switch condition reproducible, while also defining its scope:
it is a joint source, distance, and boundary intervention.

Raw models and training-free adaptations are evaluated on the full corpus.
Post-training uses one approximately 80/20 item split within each scenario,
including 320/80 speaker-switch pairs. Only the training partition enters SFT or
GRPO, and all reported post-training addressedness scores use the disjoint
held-out partition. This protocol establishes item-level separation; it does not
establish held-out voice, template-family, or tool-family generalization.

\subsection{Evaluation Contract}
Every system receives the same tool inventory and instruction to emit exactly one
of three action forms. The canonical mute output is \texttt{[Mute]}. A tool action
is a single structured object,
\texttt{<|TOOL|>\{\{"name": name, "params": \{...\}\}\}</|TOOL|>}, and an
answer contains neither a canonical tool call nor \mute. We use greedy decoding
throughout. Model-specific wrappers preserve required chat templates and response
terminators, but evaluation always applies the same normalized parser and target
action.

We report scenario-specific metrics jointly. Speaker-switch evaluation combines
switch-condition mute rate with target-tool selection in the same-speaker and
text-only controls, preventing an always-mute policy from scoring well. Self-talk
uses mute rate, and side-talk measures whether the model acts only on the
assistant-directed utterance. A correct tool output must be parseable and match
the target tool name. Arguments remain in prediction records but are outside the
\bench score, so the metric measures tool-name selection rather than complete
executable-call accuracy.

Side-talk tool items use the canonical parser. For free responses, a fixed
Qwen3.5-35B-A3B judge assigns one of four labels: assistant-directed,
bystander-directed, both, or neither. Only assistant-directed alone is correct.
The same system instruction, decoding configuration, parser, judge, prediction
schema, and summary procedure are frozen for each comparison block. The current
protocol has no reported human-agreement estimate for the judge; this affects the
free-response subset rather than the rule-scored mute and tool decisions.
Appendix~\ref{app:protocol} records the result artifacts and scorer provenance.

\section{Off-the-Shelf Action Policies}
\label{sec:findings}

\subsection{Setup}
We evaluate six raw Audio LLMs: Qwen3-Omni-30B\cite{xu2025qwen3},
Nemotron-3-Nano-Omni-30B\cite{deshmukh2026nemotron}, Gemini-3.8-flash\cite{doshi2026gemini38flash},
Kimi-Audio-7B~\citep{kimiteam2025kimiaudio},
Step-Audio-R1.1~\citep{tian2025stepaudior1}, and Audio Flamingo
3~\citep{ghosh2026audio}. We also test local, method-style adaptations
of AURA, Thinking with Sound (TwS), and SHANKS on the Qwen backbone
\citep{maben2025aura,xiong2025tws,chiang2025shanks}. These implementations probe
inference-time prompting, acoustic tools, and chunked reasoning; they are not
claimed as official reproductions. All systems use temperature zero, the same
action instruction, and the same scorer. Model-specific normalization maps
native outputs to the canonical parser, so the raw comparison reflects both
action behavior and compatibility with the standardized interface.

\begin{table*}[t]
\centering
\small
\caption{Action-selection results on the complete \bench corpus. Values are
percentages. Speaker-switch reports switch mute rate and target-tool selection
in the near-field same-speaker and text-only positive controls. Side-talk combines canonical tool selection and
free responses scored by the fixed Qwen3.5-35B judge.}
\label{tab:raw-diagnostic}
\renewcommand{\arraystretch}{1.06}
\setlength{\tabcolsep}{4pt}
\begin{tabular*}{\textwidth}{@{\extracolsep{\fill}}lrrrrr@{}}
\toprule
Method & \shortstack{Side-talk\\correct} & \shortstack{Self-talk\\mute} &
\shortstack{Switch\\mute} & \shortstack{Same-speaker\\select.} &
\shortstack{Text-only\\select.} \\
\midrule
\multicolumn{6}{l}{\textit{Raw Audio LLMs}} \\
Qwen3-Omni & 39.0 & 57.0 & 1 & 77 & 81 \\
Nemotron-Omni & 32.7 & 28.7 & 0 & 84 & 85 \\
Gemini-3.8-flash & 32.4 & 68.6 & 7 & 80 & 71 \\
Kimi-Audio-7B & 38.2 & 30.5 & \textbf{14} & 53 & 44 \\
Step-Audio-R1.1 & 41.0 & 22.4 & 1 & \textbf{96} & 97 \\
Audio Flamingo 3 & 36.0 & 1.0 & 0 & 0 & 0 \\
\midrule
\multicolumn{6}{l}{\textit{Training-free adaptations}} \\
Qwen + AURA adaptation & \textbf{46.6} & 63.2 & 0 & 95 & \textbf{99} \\
Qwen + TwS adaptation & 23.0 & \textbf{85.7} & 6 & 90 & 82 \\
Qwen + SHANKS adaptation & 17.5 & 79.8 & 1 & 1 & 81 \\
\bottomrule
\end{tabular*}
\end{table*}

\paragraph{Tool recognition does not imply acoustic-context gating.}
Step-Audio selects the target tool on 96\% and 97\% of same-speaker and
text-only controls, respectively, yet mutes only 1\% of switched commands. Kimi
has the highest raw switch mute rate at 14\%, but selects the target tool on only
53\% and 44\% of the two controls. These paired measurements expose failure to
change the action under a source-and-scene shift, rather than a general inability
to recognize tools.

\paragraph{Systems fail through different shortcuts.}
Qwen preferentially acts on the first side-talk utterance, Gemini favors the
later utterance, and Step often responds to both. Audio Flamingo 3 receives zero
canonical tool-selection credit because it emits a different bracketed action
language. This format mismatch limits cross-model comparison of positive-control
scores, but it does not explain its 1\% self-talk mute rate or zero switch mute
rate. The standardized interface therefore diagnoses deployable action behavior,
not a format-invariant latent capability.

\paragraph{Inference-time adaptations change the error profile.}
AURA improves side-talk from 39.0\% to 46.6\% and self-talk from 57.0\% to
63.2\%, but mutes no switched commands. TwS raises self-talk muting to 85.7\%
while side-talk falls to 23.0\%, largely through overuse of silence. SHANKS also
raises self-talk muting, but its chunked output often violates the action grammar,
leaving 1\% same-speaker tool selection. Acoustic evidence may appear in
intermediate reasoning without controlling the final action, an action-level
counterpart to the utilization gap studied by VoxParadox.

\section{Can Acoustic-Context Gating Be Learned?}
\label{sec:method}

We use \textsc{VoxGate} as a post-training intervention rather than as evidence
for a new speaker-identification mechanism. Supervised fine-tuning is the primary
intervention; an exploratory GRPO stage tests whether paired rollouts preserve
or improve the learned action boundary. Both stages use the same action contract
as the benchmark.

\subsection{Supervised joint training}
The model outputs $y\in\{\textsc{Mute},\textsc{Tool},\textsc{Answer}\}$. We
train a LoRA adapter on a mixture of the \bench training partition and WearVox
answer, abstention, tool-use, and translation examples. The addressedness data
include side-talk attribution, self-talk mute targets, and speaker-switch
counterfactuals. In each switch pair, matched words support a tool call for the
same-source near-field condition and \mute for the bystander condition; text-only
examples retain the tool target.

We use Qwen3-Omni-30B-A3B-Instruct with LoRA rank 32 and alpha 64 on attention
$q$, $k$, $v$, and $o$ projections across 48 layers. The audio encoder and
aligner are frozen. Training uses three epochs, learning rate $10^{-4}$,
effective batch size 32, and eight H200 GPUs.

For the cue-balanced SFT control, we replace only 1,212 speaker-switch audio training
slots in the 5,748-row mixture. Other task proportions, 540 optimization steps,
and the LoRA configuration remain fixed. Within both tool and \mute labels,
near/far rendering is crossed with 0/600~ms separation and balanced across the
four combinations. The mixture includes far-field authorized-source/tool rows, making this
a cue-control experiment rather than a proximity-gate training protocol.
Appendix~\ref{app:balanced-sft} details the construction.

\subsection{Exploratory counterfactual-pair GRPO}
For a speaker-switch pair $q=(x^{+},x^{-})$, Stage II samples $K/2$ completions
from each condition and places them in one rollout group. Here $x^{+}$ is the
same-speaker condition with a tool target, and $x^{-}$ is the switched condition
with a mute target. Grouping matched text with opposite actions makes reward
normalization depend on whether the policy separates the counterfactual pair,
rather than on unrelated prompt difficulty.

Let $r(y_{q,i},x_{q,i})$ be the rule-based reward for completion $i$. We compute
\begin{equation}
A_{q,i}=\frac{r(y_{q,i},x_{q,i})-\mu_q}{\sigma_q+\epsilon},
\qquad
\mu_q=\frac{1}{K}\sum_{i=1}^{K}r(y_{q,i},x_{q,i}),
\label{eq:paired-advantage}
\end{equation}
where $\mu_q$ and $\sigma_q$ are calculated across both sides of the pair. Only
groups with nonzero reward variance contribute an update. This removes groups in
which every sampled action receives the same reward and hence provides no
within-group policy-gradient signal.

A correct mute or matching tool receives $+1$, and an incorrect action receives
$-1$; malformed tool syntax incurs an additional $-0.2$. For answer targets,
nonempty natural language receives $+0.5$, mute receives $-1$, and empty or
tool-only output receives $-0.5$. WearVox examples retain task-specific rewards
for answer, abstention, structured tool use, and translation. We track rewards by
scenario because an aggregate curve can hide a saturated positive control or a
failed mute boundary. The post-training mixture, frozen components, and decoding
contract are shared with SFT unless stated otherwise.

This stage tests whether paired optimization preserves or improves the learned
action boundary. The experiment has no unpaired-GRPO or repeated-seed control, so
it does not isolate pair-aware grouping as the cause of an SFT-to-GRPO difference.
Appendix~\ref{app:voxgate-details} records the full implementation configuration.

\section{Post-training Results}
\label{sec:exp}

The base, supervised, and GRPO models use the same system instruction, greedy
decoding, canonical action parser, and fixed Qwen3.5-35B-A3B\cite{qwen3.5} side-talk judge.
All addressedness results use the disjoint held-out partition. Task-only SFT and
GRPO controls use WearVox without \bench examples. Downstream evaluation follows
the fixed 384-example WearVox protocol.

\begin{table*}[t]
\centering
\small
\caption{Addressedness results on the fixed 20\% \bench test split. Values are
percentages. Switch mute is reported with near-field same-speaker and text-only target-tool
selection to expose all-mute behavior.}
\label{tab:voxgate-main}
\renewcommand{\arraystretch}{1.08}
\begin{tabular*}{\textwidth}{@{\extracolsep{\fill}}lrrrrr@{}}
\toprule
Model or training setting
& \shortstack{Self-talk\\mute}
& \shortstack{Side-talk\\correct}
& \shortstack{Same-speaker\\select.}
& \shortstack{Switch\\mute}
& \shortstack{Text-only\\select.} \\
\midrule
Base Qwen & 50.0 & 53.2 & 81.3 & 0.0 & 87.5 \\
\textsc{VoxGate} supervised training & 52.0 & 68.4 & \textbf{100.0} & 91.3 & \textbf{100.0} \\
\textsc{VoxGate} + GRPO & \textbf{60.0} & \textbf{70.9} & \textbf{100.0} & \textbf{92.5} & \textbf{100.0} \\
\bottomrule
\end{tabular*}
\end{table*}

\FloatBarrier
\paragraph{Supervised training accounts for most of the switch result.}
It mutes 73 of 80 held-out switched commands while selecting the target tool for
every near-field same-speaker and text-only control. GRPO changes the switch result to 74 of
80, raises self-talk muting from 52.0\% to 60.0\%, and raises side-talk accuracy
from 68.4\% to 70.9\%. These are descriptive differences from one training run.
They show that GRPO preserves the SFT gate, but do not establish that pair-aware
grouping caused the improvement.

\paragraph{The aggregate switch score combines several cues.}
Table~\ref{tab:switch-validity-main} factorizes trigger source, distance, and the
600~ms boundary on the same 80 commands. With near-field rendering and a fixed
gap, changing only the trigger source raises mute rate from 1.25\% to 50.0\% for
SFT and to 53.75\% for GRPO. Changing a same-speaker trigger from near-field to
far-field raises muting to 60.0\% for both models, appropriate under the
proximity rule. The gap alone changes same-speaker near-field muting by only 1.25
points. The 91.3--92.5\% main switch result therefore combines source and
distance dependence.

\begin{table}[!ht]
\centering
\small
\caption{Mute rates (\%) on the original 80-command cue validity WAVs. Same/near
targets a tool; same/far and different-speaker conditions target \mute. The last
row is the original switch condition.}
\label{tab:switch-validity-main}
\setlength{\tabcolsep}{4pt}
\begin{tabular}{lllrrr}
\toprule
Source & Field & Gap & Base & SFT & GRPO \\
\midrule
Same & Near & 0 & 7.50 & 0.00 & 0.00 \\
Same & Near & 600 & 5.00 & 1.25 & 1.25 \\
Same & Far & 600 & 6.25 & 60.00 & 60.00 \\
Different & Near & 0 & 6.25 & 35.00 & 31.25 \\
Different & Near & 600 & 7.50 & 50.00 & 53.75 \\
Different & Far & 600 & 0.00 & 91.25 & 92.50 \\
\bottomrule
\end{tabular}
\end{table}

\subsection{Cue-balanced SFT as a cue control}
We call the supervised model in Tables~\ref{tab:voxgate-main} and
\ref{tab:switch-validity-main} original SFT; cue-balanced SFT is a separate
SFT-only run, with no subsequent GRPO stage. The original SFT learned from
switch examples in which bystander speech,
far-field rendering, and a 600~ms gap occurred together. We retrain with the
same budget after balancing distance and gap within both action labels, then
test all eight source, distance, and gap combinations on 80 held-out commands.

\begin{table}[!ht]
\centering
\small
\caption{Target-action accuracy (\%) under the proximity rule in the
scene-consistent eight-condition test, original SFT $\rightarrow$ cue-balanced
SFT. Each cell has $n=80$; equal-weight accuracy is 39.22\% $\rightarrow$
66.88\%. Same-speaker triggers target a tool when near and \mute when far;
different-speaker triggers target \mute. Complete mute/tool rates appear in
Appendix~\ref{app:balanced-sft}.}
\label{tab:balanced-cues-main}
\setlength{\tabcolsep}{4pt}
\begin{tabular}{lrr}
\toprule
Field, gap & Same: Tool near, \mute far & Different: \mute \\
\midrule
Near, 0~ms & 97.50 $\rightarrow$ 97.50 & 18.75 $\rightarrow$ 68.75 \\
Near, 600~ms & 100.00 $\rightarrow$ 98.75 & 18.75 $\rightarrow$ 85.00 \\
Far, 0~ms & 2.50 $\rightarrow$ 3.75 & 23.75 $\rightarrow$ 80.00 \\
Far, 600~ms & 6.25 $\rightarrow$ 5.00 & 46.25 $\rightarrow$ 96.25 \\
\bottomrule
\end{tabular}
\end{table}

\FloatBarrier
Direct retesting on the original validity WAVs raises same/far/600 muting from
60.00\% to 90.00\% and different/near muting from 35.00\%/50.00\% to
66.25\%/82.50\% at 0/600~ms (Appendix~\ref{app:switch-validity},
Table~\ref{tab:balanced-on-validity-wavs}). The eight-condition study instead
uses scene-consistent far-field audio, rendering the wearer's context and
trigger together; it yields only 3.75\%/5.00\% muting of same/far triggers.
These are different WAVs, so distance rejection is scene dependent.


\subsection{Downstream Task Retention}
The addressedness objective should not improve muting by erasing the model's
ability to answer, use tools, or translate. We therefore evaluate the
post-trained models on the fixed 384-example WearVox protocol. It contains 59
closed-book answer examples, 55 grounded-answer examples, 58 abstention examples,
112 tool-use examples, and 100 live-translation examples. The $A$ score aggregates
the two answer subsets.

\begin{table}[!ht]
\centering
\scriptsize
\caption{Held-out WearVox task retention for the original SFT/GRPO training
route (cue-balanced SFT is reported separately above). $A$ aggregates closed-book and grounded
answers; $B$, $C$, and $D$ denote abstention, tool use, and live translation.
Counts are shown in the column headers. $D$ is the mean judged translation
score ($\times 100$); Overall thresholds each translation item at 0.85 before
aggregating all 384 binary outcomes.}
\label{tab:wearvox-main}
\renewcommand{\arraystretch}{1.06}
\setlength{\tabcolsep}{4pt}
\begin{tabular}{lrrrrr}
\toprule
Model & $A$ ($n=114$) & $B$ ($n=58$) & $C$ ($n=112$) & $D$ ($n=100$) & Overall ($n=384$) \\
\midrule
Base Qwen & 28.11 & 96.60 & 55.40 & 65.30 & 45.57 \\
\multicolumn{6}{l}{\textit{WearVox-only post-training}} \\
SFT & 33.33 & 94.83 & 85.71 & 84.80 & 63.28 \\
SFT + GRPO & 42.98 & 98.28 & 86.61 & 83.43 & 67.71 \\
\multicolumn{6}{l}{\textit{Joint VoxGate post-training}} \\
\textsc{VoxGate} supervised training & 37.71 & 91.40 & 90.20 & 89.50 & 72.14 \\
\textsc{VoxGate} + GRPO & 42.10 & \textbf{100.00} & \textbf{92.00} & \textbf{89.70} & \textbf{76.30} \\
\bottomrule
\end{tabular}
\end{table}

Joint training retains downstream task performance. Task-only SFT and GRPO reach
63.28\% and 67.71\% overall, while joint supervised training and joint GRPO reach
72.14\% and 76.30\%. The joint models improve tool use and translation relative
to task-only controls; joint GRPO also attains 100.00\% abstention accuracy. The
answer aggregate remains lower than the other task families for every system,
including the base model, and the joint GRPO answer score is close to the
task-only GRPO score. These results show that the learned benchmark gate does not
come from a general collapse of retained tasks. WearVox retention does not,
however, measure transfer to natural addressedness interactions because its
downstream categories and scorer serve a different evaluation purpose.

\FloatBarrier
\section{Discussion}
\label{sec:discussion}

\paragraph{Action selection is distinct from command recognition.}
The strongest positive-control results coexist with near-zero switch muting for
off-the-shelf models. Under the standardized interface, an Audio LLM can recover
a command and select its tool without using acoustic context to decide whether
the action should execute. \bench makes this gap observable by pairing positive
and negative actions with the same specified words.

\paragraph{The learned gate uses source and proximity cues.}
At fixed near-field rendering and a 600~ms gap, changing the trigger source
raises muting by 48.75 points for SFT and 52.50 for GRPO. At fixed source and
gap, far-field rendering raises it by 58.75 points for both. Under the proximity
rule, the original SFT's 60\% same/far muting is appropriate, while its
35--50\% different/near muting leaves many bystander triggers executable. The
gap amplifies source changes but is not a standalone mute rule.

\paragraph{Paired reporting separates recognition from policy errors.}
A mute score alone rewards a degenerate policy that never acts, while tool
accuracy alone rewards a policy that executes every recognized command. The
near-field same-speaker and text-only controls expose false rejection;
different-speaker conditions expose false execution. The factorized study
changes one cue at a time while preserving the command and matching-tool identity, revealing
near-field bystander errors that an aggregate score would conceal.

\paragraph{Implications for voice-agent design.}
An action gate needs both proximity and source evidence: distance alone cannot
reject a nearby bystander. Evaluation should report false execution and false
rejection separately.

\paragraph{The scenario families test different information pathways.}
Side-talk tests recipient attribution, self-talk tests whether discourse cancels
command force, and speaker-switch changes source and scene while holding words
fixed. Performance on one family does not imply performance on the others, so we
report them separately.

\paragraph{What post-training establishes.}
Supervised training learns the benchmark's conditional action mapping while
preserving near-field and text-only tool controls and WearVox tasks. The cue
study shows why an aggregate action reversal cannot identify the rule: source
and distance can each change the decision.

\paragraph{Cue balancing improves near-bystander suppression but is scene dependent.}
The equal-budget SFT control crosses distance and gap within both training
labels, including far-field wearer/tool rows; it probes cue sensitivity rather
than training the proximity policy. On the original validity WAVs, it raises
different/near muting
from 35.00\%/50.00\% to 66.25\%/82.50\% and same/far muting from 60.00\% to
90.00\%. On scene-consistent eight-condition audio, proximity-rule accuracy
rises from 39.22\% to 66.88\%, but same/far muting remains only 3.75\%/5.00\%.
Original speaker-switch muting falls from 73/80 to 70/80, and WearVox tool-call
accuracy falls from 90.18\% to 83.04\%. The matched-WAV gains do not establish
a reliable proximity gate across acoustic scenes.

\paragraph{Current evidence boundaries.}
Results use one split and one run per setting. The balanced test varies wearer
voice but not room, template, natural bystander speech, or open-set speakers.
Side-talk uses an LLM judge without reported human agreement; \bench tool
accuracy checks names rather than arguments.

\section{Conclusion}

\bench shows that Audio LLMs often recognize spoken commands without reliably
choosing the intended action across 1,018 controlled items. Supervised training
learns much of this benchmark action boundary, while an exploratory GRPO stage
also adds small gains in one run. Factorized controls identify an independent
source-change effect, but sensitivity to far-field rendering varies across
acoustic scenes. Cue-balanced SFT improves near-bystander suppression without
establishing a reliable cross-scene proximity gate. These results highlight a
gap between understanding what was said and deciding whether acoustic and
conversational context warrants action, and motivate evaluating voice agents at
the level of executable action rather than command recognition alone.

\clearpage

\bibliography{references}
\bibliographystyle{iclr2027_conference}

\appendix

\section{Evaluation Contract and Reproducibility Record}
\label{app:protocol}

Each result is tied to a frozen benchmark manifest, system instruction, decoding
configuration, action parser, side-talk judge, prediction file, and summary
file. Raw Audio LLMs and training-free adaptations are evaluated on all retained
\bench items. Post-trained systems are trained on an approximately 80\% partition
and reported only on the disjoint held-out partition. We use greedy decoding throughout.
The full corpus contains 395 side-talk recordings, 223 self-talk recordings,
and 400 speaker-switch pairs. Each speaker-switch pair has single, switch, and
text-only conditions; the held-out post-training speaker-switch set therefore
contains 80 cases per condition.

\begin{table*}[!h]
\centering
\small
\caption{Result blocks and evaluation records. ``Full'' denotes the complete
retained \bench corpus, while ``held-out'' denotes the fixed post-training test
partition.}
\label{tab:appendix-provenance}
\renewcommand{\arraystretch}{1.08}
\scriptsize
\begin{tabular}{@{}p{0.14\textwidth}p{0.16\textwidth}p{0.08\textwidth}p{0.08\textwidth}p{0.295\textwidth}@{}}
\toprule
Result block & Systems & Split & Decoding & Scoring record \\
\midrule
Raw Audio LLMs & Six native systems & Full & Greedy & Canonical parser; Qwen3.5 side-talk judge \\
Training-free methods & AURA, TwS, SHANKS & Full & Greedy & Same action contract and judge \\
VoxGate gating & Base, SFT, GRPO & Held-out & Greedy & Same instruction, parser, and judge \\
WearVox retention & Base and post-training controls & 384 items & Greedy & Official A/B/C/D scorer and translation judge \\
\bottomrule
\end{tabular}
\normalsize
\end{table*}

\section{Action Grammar and Scoring}
\label{app:scoring}

The canonical mute output is \texttt{[Mute]}. A canonical tool action is a
single structured object,
\texttt{<|TOOL|>\{\{"name": \textit{name}, "params": \{...\}\}\}</|TOOL|>}.
A natural-language answer contains neither a canonical tool call nor \mute. The
\bench parser normalizes the mute token and checks that a tool output contains
one parseable object with the target tool name. Tool arguments are retained in
prediction logs but are not included in \bench tool-selection accuracy; they are
evaluated by the downstream WearVox tool-use task.

For side-talk free-response examples, the fixed judge selects one of four labels:
assistant-directed utterance, bystander-directed utterance, both, or neither.
Only the first label is correct. Speaker-switch mute is always interpreted with
same-speaker and text-only tool-selection controls, since a system that always
mutes cannot be a valid action gate.

\section{Complete Results for Raw and Training-Free Systems}
\label{app:raw-results}

Table~\ref{tab:raw-diagnostic} reports the compact comparison. Table~\ref{tab:appendix-raw-full}
adds the scenario denominators and separates the raw and training-free blocks.
The values are the same measurements as the main table, rather than a second
selection pass. Side-talk combines tool items scored by the canonical parser and
response items scored by the fixed side-talk judge.

\begin{table*}[t]
\centering
\small
\caption{Complete full-corpus raw and training-free results. The corpus contains
395 side-talk recordings, 223 self-talk recordings, and 400 speaker-switch
pairs. All values are percentages.}
\label{tab:appendix-raw-full}
\renewcommand{\arraystretch}{1.06}
\begin{tabular*}{\textwidth}{@{\extracolsep{\fill}}lrrrrr@{}}
\toprule
Method & Side-talk & Self-talk mute & Switch mute & Single select. & Text-only select. \\
\midrule
\multicolumn{6}{l}{\textit{Raw Audio LLMs}} \\
Qwen3-Omni & 39.0 & 57.0 & 1 & 77 & 81 \\
Nemotron-Omni & 32.7 & 28.7 & 0 & 84 & 85 \\
Gemini-3.8-flash & 32.4 & 68.6 & 7 & 80 & 71 \\
Kimi-Audio-7B & 38.2 & 30.5 & 14 & 53 & 44 \\
Step-Audio-R1.1 & 41.0 & 22.4 & 1 & 96 & 97 \\
Audio Flamingo 3 & 36.0 & 1.0 & 0 & 0 & 0 \\
\midrule
\multicolumn{6}{l}{\textit{Training-free adaptations}} \\
Qwen + AURA & 46.6 & 63.2 & 0 & 95 & 99 \\
Qwen + TwS & 23.0 & 85.7 & 6 & 90 & 82 \\
Qwen + SHANKS & 17.5 & 79.8 & 1 & 1 & 81 \\
\bottomrule
\end{tabular*}
\end{table*}

The three training-free methods use no weight update. AURA adds a ReAct-style
attribution step before final action selection. TwS supplies a bounded acoustic
analysis protocol with local voice, pitch, energy, and spectral tools. SHANKS
uses chunked listening and intermediate notes. These are local implementations
of the respective method ideas and are not claimed as official full-pipeline
reproductions.

Kimi-Audio receives the system instruction in its first user message because its
native template does not retain a system role. Step-Audio uses its required
response terminators and long-audio handling. Audio Flamingo~3 emits a native
bracketed action language, which does not satisfy the canonical tool contract.
This explains its strict tool-selection zeros, but not its addressedness errors:
it mutes only 1\% of self-talk items and no switched commands.

\section{VoxGate Implementation Details}
\label{app:voxgate-details}

\begin{table}[t]
\centering
\small
\caption{VoxGate implementation configuration.}
\label{tab:appendix-voxgate-config}
\renewcommand{\arraystretch}{1.08}
\begin{tabular}{p{0.40\linewidth}p{0.52\linewidth}}
\toprule
Component & Configuration \\
\midrule
Base model & Qwen3-Omni-30B-A3B-Instruct \\
SFT adaptation & LoRA on all 48-layer attention $q,k,v,o$ projections; rank 32, alpha 64 \\
Frozen modules & Audio encoder and aligner \\
SFT optimization & Three epochs, learning rate $10^{-4}$, effective batch size 32, eight H200 GPUs \\
SFT mixture & WearVox answer, abstention, tool-use, and translation data plus the approximately 80\% \bench training partition \\
GRPO mixture & The same downstream task mixture and addressedness examples; pair-aware grouping for speaker-switch \\
Rollouts & $K/2$ completions for the act and mute sides of each speaker-switch pair \\
Action rewards & Mute, target tool, answer, and task-specific WearVox rewards \\
Evaluation & Greedy decoding; fixed instruction, parser, and side-talk judge \\
\bottomrule
\end{tabular}
\end{table}

Equation~\ref{eq:paired-advantage} defines the paired normalized advantage, and
Section~\ref{sec:method} specifies the action rewards. The implementation records
per-family reward distributions and the number of nonzero-variance rollout
groups so that addressedness updates can be distinguished from retained-task
updates.

\section{Downstream WearVox Evaluation}
\label{app:wearvox}

Table~\ref{tab:wearvox-main} reports the complete 384-example WearVox breakdown,
including task denominators. Although WearVox-only and joint-model outputs are
archived separately, all Table~\ref{tab:wearvox-main} rows were evaluated on the
same 384 items with identical task prompts, judge model and prompt, and A/B/C/D
scoring code.

\section{Construction and Quality Controls}
\label{app:quality}

Side-talk scripts are balanced over utterance order and tool-like versus
conversational content, producing eight order/content-pairing configurations.
Self-talk items place a command core in discourse frames including planning,
regret, quotation, sarcasm, and rhetorical questions. Recordings are converted
to 16~kHz mono PCM WAV and retained only after annotation, label, and
script-audio consistency checks.

For speaker-switch, 400 cases are balanced across ten consequential tools
(40 cases per tool). Eight distinct voices are used to synthesize speaker
changes. The single condition renders source A for both context and trigger.
The switch condition renders source A for context, inserts 600~ms of silence,
and renders the trigger from source B. The B trigger receives fixed far-field
room rendering. Audio is normalized to
$-20$~dBFS per segment with 5~ms fades and 16~kHz mono PCM output. The synthesis
manifest records voice, duration, normalization, peak, boundary, and room
configuration for each case.

\begin{table}[t]
\centering
\small
\caption{Speaker-switch construction controls.}
\label{tab:appendix-switch-controls}
\renewcommand{\arraystretch}{1.08}
\begin{tabular}{p{0.38\linewidth}p{0.54\linewidth}}
\toprule
Variable & Controlled construction \\
\midrule
Specified text & Matched context and trigger across single, switch, and text-only conditions \\
Action target & Tool in single/text-only; \mute in switch \\
Voice synthesis & Eight distinct voices simulate speaker changes \\
Temporal boundary & 600~ms silence before the switch trigger \\
Acoustic scene & Far-field room rendering on the B trigger only \\
Coverage & 400 cases, 10 tools, 40 cases per tool \\
\bottomrule
\end{tabular}
\end{table}

\section{Speaker-Switch Cue Validity Study}
\label{app:switch-validity}

We construct an evaluation-only validity set from the 80 frozen held-out
speaker-switch commands. The set is excluded from training, prompt tuning,
checkpoint selection, and model selection. For each command, we synthesize the
wearer context, wearer trigger, and bystander trigger once and reuse those source
segments across six conditions. This construction holds text, tool target, TTS
configuration, sample rate, segment-level loudness, and scoring fixed while
varying speaker relation, trigger distance, and temporal separation. Far-field
segments are normalized after room rendering so that distance does not introduce
a systematic level difference. All 480 generated files pass format, hash, peak,
gap, and segment-level loudness checks.

\begin{table}[t]
\centering
\small
\caption{Controlled conditions in the speaker-switch validity set. Each
condition contains the same 80 held-out commands. Targets follow the proximity
rule.}
\label{tab:switch-validity-design}
\renewcommand{\arraystretch}{1.06}
\begin{tabular}{llll}
\toprule
Trigger source & Trigger field & Gap & Target \\
\midrule
Same speaker & Near & 0~ms & Tool \\
Same speaker & Near & 600~ms & Tool \\
Same speaker & Far & 600~ms & \mute \\
Different speaker & Near & 0~ms & \mute \\
Different speaker & Near & 600~ms & \mute \\
Different speaker & Far & 600~ms & \mute \\
\bottomrule
\end{tabular}
\end{table}

Table~\ref{tab:switch-validity-results} reports mute and matching-tool selection
rates separately for every condition. The raw base model changes little across
the six conditions. Both post-trained models respond to source change under
near-field rendering, but they also use distance as a strong gating cue. The
600~ms boundary alone has little effect on same-speaker commands and instead
amplifies muting when the trigger source changes.

\begin{table*}[t]
\centering
\scriptsize
\caption{Speaker-switch cue validity results on the frozen 80-command test set.
Cells report mute rate (M) and matching-tool selection rate (T) in percent.
Under the proximity rule, same/near targets T and all other audio rows target M.
The final row is the original
switch condition from Table~\ref{tab:voxgate-main}; its non-target tool rate was
not retained in the reported summary and is marked with ``--''.}
\label{tab:switch-validity-results}
\renewcommand{\arraystretch}{1.08}
\begin{tabular*}{\textwidth}{@{\extracolsep{\fill}}llllrrrrrr@{}}
\toprule
Source & Field & Gap & Target & \multicolumn{2}{c}{Base Qwen} & \multicolumn{2}{c}{VoxGate SFT} & \multicolumn{2}{c}{VoxGate GRPO} \\
\cmidrule(lr){5-6}\cmidrule(lr){7-8}\cmidrule(lr){9-10}
& & & & M & T & M & T & M & T \\
\midrule
Same & Near & 0~ms & Tool & 7.50 & 65.00 & 0.00 & 100.00 & 0.00 & 100.00 \\
Same & Near & 600~ms & Tool & 5.00 & 68.75 & 1.25 & 98.75 & 1.25 & 98.75 \\
Same & Far & 600~ms & \mute & 6.25 & 73.75 & 60.00 & 40.00 & 60.00 & 40.00 \\
Different & Near & 0~ms & \mute & 6.25 & 75.00 & 35.00 & 65.00 & 31.25 & 68.75 \\
Different & Near & 600~ms & \mute & 7.50 & 71.25 & 50.00 & 50.00 & 53.75 & 46.25 \\
Different & Far & 600~ms & \mute & 0.00 & -- & 91.25 & -- & 92.50 & -- \\
\bottomrule
\end{tabular*}
\end{table*}

The paired contrasts clarify the contribution of each cue. At fixed near-field
rendering and 600~ms separation, changing the trigger speaker increases mute rate
by 48.75 points for SFT and 52.50 points for GRPO. At fixed speaker identity and
600~ms separation, the far-field transform increases mute rate by 58.75 points
for both models. Removing the gap from a different near-field source reduces mute
rate by 15.0 points for SFT and 22.5 points for GRPO. In contrast, adding the gap
to the same near-field wearer changes mute rate by only 1.25 points and preserves
98.75\% target-tool selection. The original speaker-switch condition therefore
combines source, distance, and turn-boundary evidence, while the same-speaker
control shows that the temporal pause is not learned as a standalone mute rule.

\begin{table}[!ht]
\centering
\small
\caption{Cue-balanced SFT on the original six-condition validity WAVs
($n=80$ per condition), using the same prompts, temperature $0$, parser, and
scorer as the original validity study. M is \mute rate and T is matching-tool
selection rate (\%). These WAVs differ from the scene-consistent eight-condition
set in Table~\ref{tab:balanced-cues-full}.}
\label{tab:balanced-on-validity-wavs}
\setlength{\tabcolsep}{4pt}
\begin{tabular}{lllrr}
\toprule
Source & Field & Gap & M & T \\
\midrule
Same & Near & 0~ms & 2.50 & 97.50 \\
Same & Near & 600~ms & 2.50 & 97.50 \\
Same & Far & 600~ms & 90.00 & 10.00 \\
Different & Near & 0~ms & 66.25 & 33.75 \\
Different & Near & 600~ms & 82.50 & 17.50 \\
Different & Far & 600~ms & 93.75 & 5.00 \\
\bottomrule
\end{tabular}
\end{table}

\FloatBarrier

\section{Cue-Balanced SFT Control}
\label{app:balanced-sft}

The cue-balanced run changes only the 1,212 frozen-train speaker-switch audio slots in
the 5,748-row original SFT mixture: 306 wearer/tool and 906 bystander/\mute
slots. The 286 speaker-switch text-only tool rows and all other task rows remain unchanged.
The four near/far by 0/600~ms combinations receive 77, 77, 76, and 76 wearer
tool rows, and 227, 227, 226, and 226 bystander mute rows, respectively.
The mixture includes far-field authorized-source/tool examples and serves as a
cue-control experiment rather than proximity-policy training.
Training retains three epochs, 540 steps, and the LoRA settings in
Table~\ref{tab:appendix-voxgate-config}. The 320 training commands and 80 test
commands are disjoint; all eight variants of a command stay within its split.
Training wearer voices are \texttt{M000}, \texttt{M008}, \texttt{mom},
\texttt{dad}, \texttt{kid\_m}, and \texttt{grandma}; test wearer voices are
\texttt{teen\_f}, \texttt{phone\_caller}, and \texttt{tv\_host}.

Each command reuses one synthesized wearer context, wearer trigger, and
bystander trigger across its eight variants. All speech segments share the same
text, TTS checkpoint, 16~kHz mono PCM format, and segment-level loudness
normalization. Far-field room rendering and low-pass filtering affect both the
wearer context and trigger in a far-field variant; rendered voiced segments are
renormalized. The 0/600~ms contrast changes only the inserted silence. In the
earlier six-condition cue-validity study, far-field rendering was a trigger-only
intervention. Consequently, even with the same 80 command texts, the old and
new conditions are different WAVs and their cells should not be pooled or used
as matched before/after measurements.

\begin{table}[!ht]
\centering
\scriptsize
\caption{Complete eight-condition results on scene-consistent audio. Each row
contains 80 held-out commands. M is \mute rate and T is matching-tool
selection rate (\%). Same/near targets T; same/far and different-speaker rows
target M under the proximity rule.}
\label{tab:balanced-cues-full}
\setlength{\tabcolsep}{5pt}
\begin{tabular}{lllrrrr}
\toprule
Relation & Field & Gap & \multicolumn{2}{c}{Original SFT} & \multicolumn{2}{c}{Cue-balanced SFT} \\
\cmidrule(lr){4-5}\cmidrule(lr){6-7}
& & & M & T & M & T \\
\midrule
Same & Near & 0~ms & 2.50 & 97.50 & 2.50 & 97.50 \\
Same & Near & 600~ms & 0.00 & 100.00 & 1.25 & 98.75 \\
Same & Far & 0~ms & 2.50 & 97.50 & 3.75 & 96.25 \\
Same & Far & 600~ms & 6.25 & 93.75 & 5.00 & 95.00 \\
Different & Near & 0~ms & 18.75 & 81.25 & 68.75 & 31.25 \\
Different & Near & 600~ms & 18.75 & 81.25 & 85.00 & 15.00 \\
Different & Far & 0~ms & 23.75 & 76.25 & 80.00 & 20.00 \\
Different & Far & 600~ms & 46.25 & 53.75 & 96.25 & 3.75 \\
\bottomrule
\end{tabular}
\end{table}

\FloatBarrier

\end{document}